\documentclass{hep99}
\usepackage{epsfig}
\begin{document}
\newcommand{\mgg} {$M_{\gamma \gamma}$}
\newcommand{\mdip} {M_{\gamma \gamma}}
\newcommand{\epair}    {\mbox{${\mathrm e}^+{\mathrm e}^-$}}
\newcommand{\epem}{{\mathrm e}^+ {\mathrm e}^-}
\newcommand{\tptm}{{\tau}^+ {\tau}^-}
\newcommand{\mpmm}{{\mu}^+ {\mu}^-}
\newcommand{\qqbar}{{\mathrm q}\bar{\mathrm q}}
\newcommand{\nn}{\nu \bar{\nu}}
\newcommand{\nunu}{\nu \bar{\nu}}
\newcommand{\mumu}{\mu^+ \mu^-}
\newcommand{\ellell}{\ell^+ \ell^-}
\newcommand{\MZ}{M_{\mathrm Z}}
\newcommand{\Mh}{$M_{\mathrm h}$}
\newcommand{\MX} {M_{\mathrm{X}}}
\newcommand{\MY} {M_{\mathrm{Y}}}

\newcommand {\Hboson}        {{\mathrm h}^{0}}
\newcommand {\Hzero}         {${\mathrm h}^{0}$}
\newcommand {\Zboson}        {{\mathrm Z}^{0}}
\newcommand {\Zzero}         {${\mathrm Z}^{0}$}
\newcommand{\mupair}   {\mbox{$\mu^+\mu^-$}}
\newcommand{\taupair}  {\mbox{$\tau^+\tau^-$}}
\newcommand{\qpair}    {\mbox{${\mathrm q}\overline{\mathrm q}$}}
\newcommand{\ff}       {{\mathrm f} \bar{\mathrm f}}
\newcommand{\gaga}     {\gamma\gamma}
\newcommand{\WW}       {{\mathrm W}^+{\mathrm W}^-}
\newcommand {\mm}         {\mu^+ \mu^-}
\newcommand {\Zz}         {\mbox{${\mathrm{Z}^0}$}}
\newcommand{\roots}     {\sqrt{s}}
\newcommand{\Ecm}         {\mbox{$E_{\mathrm{cm}}$}}
\newcommand{\Ebeam}       {E_{\mathrm{beam}}} 
\newcommand{\ipb}         {\mbox{pb$^{-1}$}}
\newcommand{\ra}        {\rightarrow}   
\newcommand{\ov}        {\overline}   
\newcommand{\OPALColl}  {OPAL Collab.}
\def\mrm       {\mathrm}

%

\title{Searches for ``Other'' Higgs Bosons at LEP}

\author{Mark J Oreglia$^1$}

\address{$^1$ The Enrico Fermi Institute, The University of Chicago, Chicago 60637, USA\\[3pt]
E-mail: {\tt m-oreglia@uchicago.edu}}

\abstract{
Recent LEP searches for Higgs bosons in models other
than the Minimal Standard Model and the Minimal Supersymmetric
Standard Model are reviewed. Limits are presented for Higgs
bosons decaying into diphotons or invisible particles, 
and for charged Higgs bosons.
}

\maketitle


\section{Introduction}

The Minimal Standard Model (MSM) incorporating a single Higgs doublet has
one neutral Higgs boson with all decay rates specified by the theory.
The dominant production mechanism at LEP is the Bjorken process
$\epem \ra \Hboson \Zboson$;
a 95.2~GeV lower mass limit on the MSM \Hzero\ is obtained by combining
the results from the four LEP experiments analyzing data up to 
\Ecm=189~GeV~\cite{LEPMSM}.

In models employing more than one Higgs doublet or triplet fields, 
a rich spectrum of Higgs bosons occurs, with possibly large numbers
of unknown decay parameters. For instance, the Minimal Supersymetric Standard Model
(MSSM) is a two Higgs doublet model (2HDM) obtained with a particular
choice of the couplings of the Higgs fields. This model has five Higgs particles
in the form of three neutrals (one is CP odd) and a singly-charged pair.  
In more generality, there are four ways to couple the 2HDM fields to fermions and
bosons (some authors classify these as model types I, I', II, and II').

In this brief review, I present results from LEP searches for Higgs bosons decaying
in the context of models other than the MSM and MSSM. 

\section{Photonically Decaying Higgs Bosons}

In the MSM, the Higgs boson can decay into a pair of photons by means of a W-loop.
For a Higgs boson of mass 80~GeV, the diphoton branching fraction is 0.001,
hence this mode is not visible at LEP. 
However, in non-minimal models, when the topology of the theory reduces
the Higgs boson coupling strength to fermions, the diphoton mode can become
large. An example of this is the ``fermiophobic'' Higgs boson~\cite{fermiophobic} arising
in the Type-I 2HDM. In this model, all the fermionic couplings to one of
the Higgs neutrals have a factor $\cos{\alpha}/\sin{\beta}$, so that the 
appropriate choice of $\alpha$ turns off the fermionic couplings. In this theory,
the lightest neutral boson is produced in \epair\ collisions at MSM strength.

Very different theories can give rise to enhanced 
$H^{0} \rightarrow \gamma \gamma$ 
rates, so it is important
to present cross section limits in addition to model-specific Higgs boson mass limits.
The list of theories having enhanced diphoton rates includes 
the 2HDM,
the Higgs Triplet model,
top-quark condensate models,
models with extra dimensions,
models with anomalous couplings,
the hypercharge axion,
etc.

\begin{figure}[!h]
\begin{center}
\vspace*{0.8cm}
\resizebox{\linewidth}{!}{\includegraphics{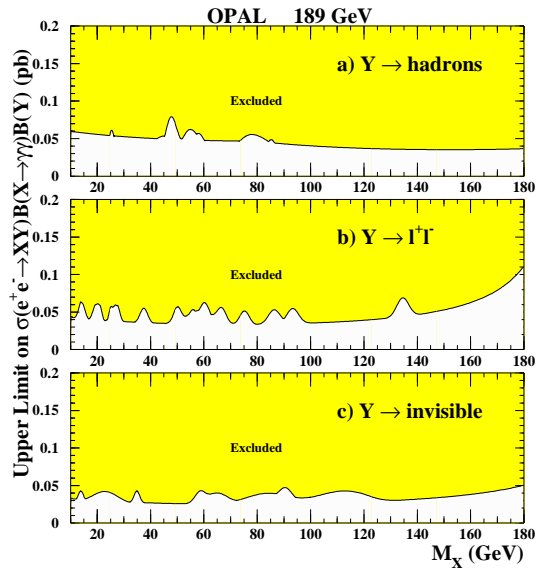} }
\end{center}
\caption{
                 95\% confidence level upper limit on
                 $\sigma(\epem\ra {\rm XY}) \times B(\mrm X \ra \gaga) \times B(Y)$
                 for the case 
                 where: a) Y decays hadronically, b) Y decays into any charged
                 lepton pair and c) Y decays invisibly. The limits
                 for each $\MX$ assume the smallest efficiency as a function of
                 $\MY$ such that $10 < \MY < 180$~GeV and that
                 $\MX + \MY > M_{\mrm Z}$.
\label{opalsig}}
\end{figure}

\begin{figure}[p]
\begin{center}
\vspace*{0.8cm}
\resizebox{\linewidth}{!}{\includegraphics{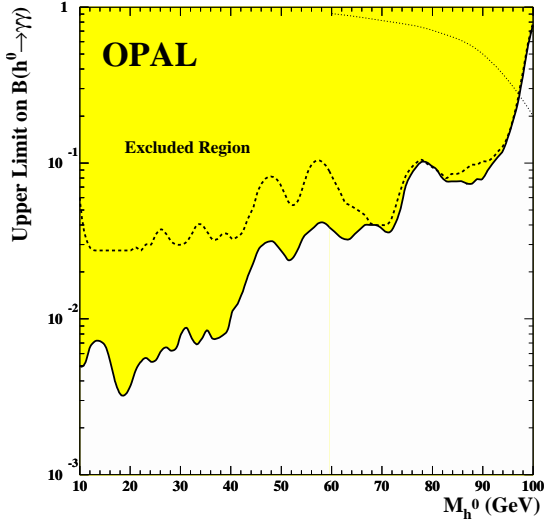} }
\end{center}
\caption{
                 95\% confidence level upper limit on the branching fraction
                 $B$($\Hboson \ra \gaga$)
                 for a Standard Model Higgs boson production rate. 
                 The shaded region, obtained with all LEP energies, is excluded;
                 the dashed line shows the limit obtained with the 189~GeV data only. 
                 The dotted line
                 is the predicted $B$($\Hboson \ra \gaga$) assuming
                 $B$($\Hboson \ra \mrm f \bar{\mrm f}$)=0. 
                 The intersection of the dotted line with the exclusion curve gives a lower limit of
                 96.2~GeV for the fermiophobic Higgs model. 
\label{opalbrg}}
\end{figure}

Figure~\ref{opalsig} shows the 95\% CL upper limits on the production cross sections
for $\sigma(\epem\ra {\rm XY}) \times B(\mrm X \ra \gaga) \times B(Y)$
obtained by OPAL~\cite{OPALgg} 
with data from \Ecm\ up to 189~GeV;
here, X can be a fermiophobic Higgs Boson and Y can be a scalar
or vector particle.
Factoring out the SM Higgs boson production cross section,
OPAL obtains upper limits on the diphoton branching ratio
(Figure~\ref{opalbrg}); similar results have been contributed to this
conference by ALEPH, DELPHI, and L3~\cite{ADLgg}.

\section{Invisibly Decaying Higgs Bosons}

In nonminimal models, the Higgs boson could decay into undetected particles
such as a pair of SUSY particles.
The Bjorken production mechanism allows searching for this mode by tagging
the recoil \Zzero. Backgrounds to this search arise primarilly from
4-fermion and WW processes.

\begin{figure}[p]
\begin{center}
\vspace*{0.8cm}
\resizebox{\linewidth}{!}{\includegraphics{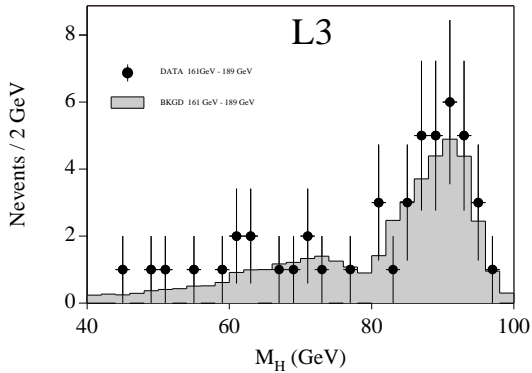} }
\end{center}
\caption{
L3 candidate events for invisible Higgs decays
in the data having \Ecm=161--189~GeV.
\label{L3inv}}
\end{figure}

\begin{figure}[p]
\begin{center}
\vspace*{0.8cm}
\resizebox{\linewidth}{!}{\includegraphics{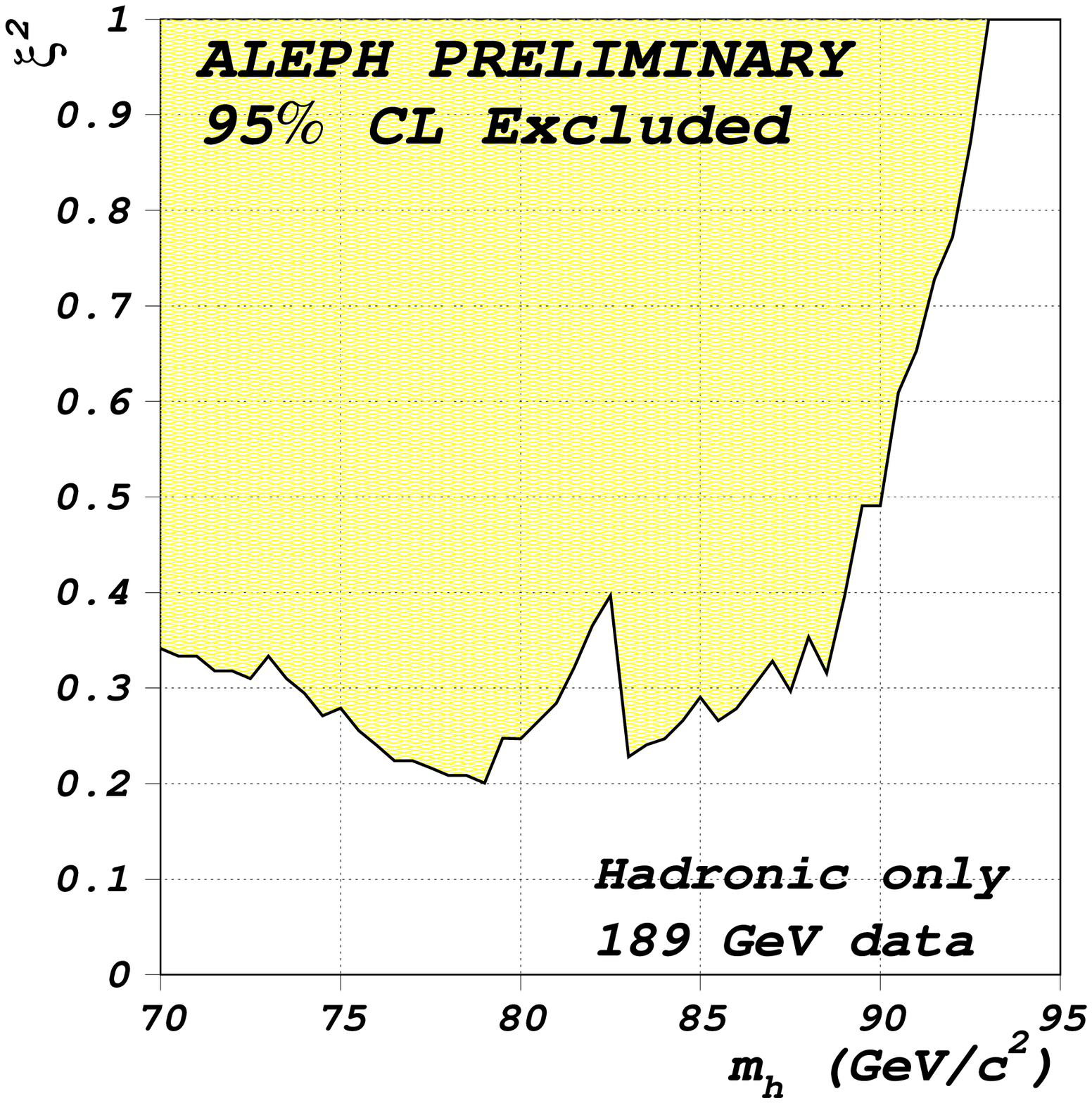} }
\end{center}
\caption{ 
Excluded region of \Mh\ vs $\xi^{2}$ for 
$\epem \ra \Hboson \Zboson, \Zboson \ra hadrons$, 
with \Hzero decaying invisibly.
\label{Ainv}}
\end{figure}

The search results may be intrepreted by assuming that the invisibly-decaying
Higgs boson is produced at the MSM rate modified by the factor $\xi$.
All the LEP experiments have presented search limits to this conference~\cite{ADLOinv}.
The L3 plot of candidate events is shown in Figure~\ref{L3inv}.  
The ALEPH limits on $\xi$ are shown in Figure~\ref{Ainv}.

\section{Charged Higgs Bosons}

Charged Higgs bosons can be pair-produced at LEP
($\epem \ra H^{+} H^{-}$). 
Models giving rise to singly-charged Higgs bosons are
the 2HDM (including the MSSM), triplet models, and
models with other extended Higgs sectors. 

In the 2HDM, the pair-production rate is specified,
however the $H^{\pm}$ decay couplings are not. 
The searches currently assume that   
$H^{\pm} \ra c \bar{s}$ (``hadron mode'') and
$H^{\pm} \ra \tau \nu_{\tau}$ (``lepton mode'')
are the dominant decays,
and therefore
BR($H^{\pm} \ra \tau \nu_{\tau}$)
is treated as a parameter of the theory.
Akeroyd~\cite{Akeroyd} suggests that $H^{\pm} \ra W^{*}A^{0}$
is an important search channel for triplet models
(and for 2HDM if $H^\pm$ is very massive).

\begin{figure}[p]
\begin{center}
\vspace*{0.8cm}
\resizebox{\linewidth}{!}{\includegraphics{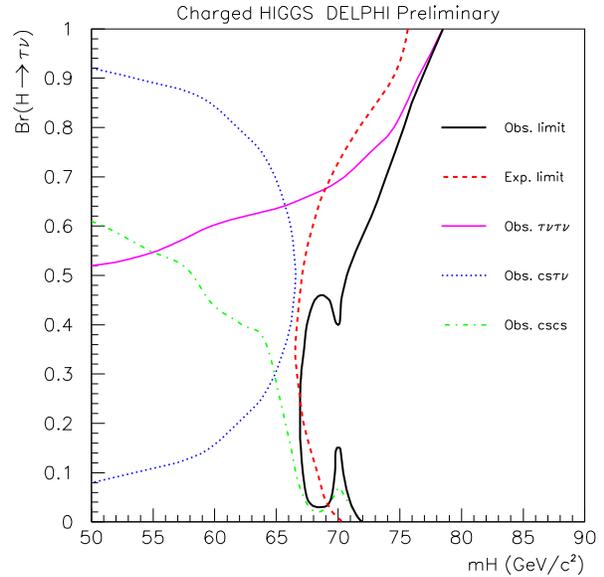} }
\end{center}
\caption{
95\% CL observed and expected exclusion regions for $H^\pm$
obtained from a combination of the search results in the hadronic, leptonic, 
and mixed channels. The data are from \Ecm=183 and 189~GeV.
\label{Dhpm}}
\end{figure}

All the LEP collaborations have submitted search updates to this 
conference~\cite{ADLOhpm}.
The search results for \Ecm\ up to 189~GeV are summarized in
Table~\ref{T1}, where the mass limit shown is the lowest
value of the 95\% CL lower bound for any value of 
BR($H^{\pm} \ra \tau \nu_{\tau}$).
The exclusion regions for the various modes in the
DELPHI analysis are shown in Figure~\ref{Dhpm}.
Also, for this conference, the LEP Higgs Working Group has
combined the 189~GeV results from the four experiments,
obtaining a lower mass bound of 77.3~GeV~\cite{LEPMSM}.

\begin{table}[!h]
\begin{center}
\caption
{LEP 95\% CL lower mass limits for charged Higgs bosons.
The expected limit is in parentheses.
\label{T1}}
\begin{tabular}{cccc} 
\br
Mode:  & Data  & Background  & Limit (GeV) \\ 
\mr
ALEPH  &       &             & 65.5 (69.5) \\
hadron & 263   & 295.4       & \\
lepton & 20    & 15.5        & \\
mixed  & 19    & 22.6        & \\
\mr
DELPHI &       &             & 66.9 (66.5) \\
hadron & 145   & 141.3       & \\
lepton & 15    & 15.8        & \\
mixed  & 55    & 55.9        & \\
\mr
L3     &       &             & 67.5 (70.2) \\
hadron & 335   & 359.4       & \\
lepton & 30    & 32.5        & \\
mixed  & 134   & 132.0       & \\
\mr
OPAL   &       &             & 68.7 (68.5) \\
hadron & 156   & 153.8       & \\
lepton & 31    & 26.2        & \\
mixed  & 65    & 60.1        & \\
\br
\end{tabular}
\end{center}
\end{table}
 


\end{document}